\documentclass[epsf,useAMS,usenatbib,usegraphicx]{mn2e} \usepackage{epsf}

\newcommand{\vsini}{$v \sin i$}
\usepackage[usenames]{color}
\newcommand{\changeb}{}

\def\teff{\hbox{$\,T_{\rm eff}$}}
\def\kms{\hbox{$\,{\rm km}\,{\rm s}^{-1}$}}
\def\ms{\hbox{$\,{\rm m}\,{\rm s}^{-1}$}}

\def\farcs{\hbox{$.\!\!^{\prime\prime}$}}

\def\logg{\hbox{$\log$\,g}}
\def\halpha{\hbox{${\rm H}\alpha$}}

\def\teff{\hbox{$\,T_{\rm eff}$}}

\def\kms{\hbox{$\,{\rm km}\,{\rm s}^{-1}$}}

\title [Spectroscopic nature of HD\,151878]{On the spectroscopic nature of the 
cool evolved Am star HD\,151878\thanks{Based on observations collected at the 
Nordic Optical Telescope as part of programme 36-418.}}

\author[Freyhammer et al.] {L.M.~Freyhammer$^{1}$ V.G.~Elkin$^{1}$ and 
D.W.~Kurtz$^{1}$\\
$^{1}$Centre for Astrophysics, University of Central Lancashire, Preston PR1 2HE, 
UK\\ }

\begin{document}

\maketitle

\begin{abstract}
Recently, \citet{tiwarietal07} detected the bright component of the visual binary 
HD\,151878 to exhibit rapid photometric oscillations through a Johnson $B$ filter 
with a period of 6\,min (2.78\,mHz) and a high, modulated amplitude up to 
22\,mmag peak-to-peak, making this star by far the highest amplitude roAp star 
known. As a new roAp star, HD\,151878 is of additional particular interest as a 
scarce example of the class in the northern sky, and only the second known case of 
an evolved roAp star -- the other being HD\,116114. We used the {\it FIES} 
spectrograph at the Nordic Optical Telescope to obtain high time resolution 
spectra at high dispersion to attempt to verify the rapid oscillations. We show 
here that the star at this epoch is spectroscopically stable to rapid oscillations 
of no more than a few tens of \ms. The high-resolution spectra furthermore show 
the star to be of type Am rather than Ap and we show the star lacks most of 
the known characteristics for rapidly oscillating Ap stars. We conclude that this 
is an Am star that does not pulsate with a 6-min period. The original discovery of 
pulsation is likely to be an instrumental artefact. 

\end{abstract}

\begin{keywords} Stars: oscillations -- stars: variables -- stars: individual 
(HD\,151878) -- stars: magnetic. \end{keywords}

\section{Introduction}

\subsection{Rapidly oscillating Ap stars}

After the discovery of low amplitude pulsation for the bright and well-studied 
cool, magnetic chemically peculiar A (Ap) star $\beta$\,CrB 
\citep{hatzesetal04,kurtzetal07}, the question arose whether all cool Ap stars 
(with effective temperatures below $T_{\rm eff}\sim 8200$\,K) are rapid 
oscillators. This still-unanswered question may be a pivotal point for theoretical 
modelling and understanding of the driving mechanism in rapidly oscillating (roAp) 
stars. 

Ap stars are characterised by overabundances of Sr, Cr, Si, Mn and rare earth 
elements. Many searches for rapid oscillations in Ap stars have been made. 
\citet{elkinetal08a} found that 25 known roAp stars that exhibit pulsations 
photometrically also show rapid radial velocity variations for the corresponding 
pulsation periods. However, several Ap stars with photometric indices typical for 
known roAp stars were tested photometrically for pulsations by 
\citet{martinezetal94} and found to be stable to high precision; these stars are 
called non-oscillating Ap stars, or noAp stars.

As seen in, e.g., the astrometric HR-diagram by \citeauthor{hubrigetal05}\, 
(\citeyear{hubrigetal05}, their figure 2) for roAp and noAp stars, the apparent 
noAp stars occupy essentially the same regions as the roAp stars. However, the 
noAp stars appear to be systematically more evolved than the roAp stars 
\citep{northetal97,handleretal99,hubrigetal00}. Nevertheless, the theoretical roAp 
instability strip \citep{cunha02} also predicts pulsations for the more evolved Ap 
stars near the terminal age main sequence, with periods in the range $16 - 
25$\,min (see Fig.\,\ref{fig:hrd} in Section~\ref{sec:params}  below). HD\,116114 
\citep{elkinetal05a} was indeed detected in this region of the HR-diagram, 
oscillating with the predicted frequency 0.79\,mHz (the lowest frequency known for 
the roAp stars). Another somewhat less-evolved roAp star, HD\,218994, has been 
recently discovered with a period of 14.2\,min, which is typical of many roAp 
stars (\citealt{gonzalez08}). Thus HD\,116114 is still the only case of a luminous 
roAp star near the terminal age main sequence; it shows extremely low radial 
velocity pulsation amplitude only for a small number of chemical elements such as 
europium and lanthanum.

\citet{freyhammeretal08a} searched for rapid pulsations among a group of 9 evolved 
Ap stars inside the roAp instability strip in the HR-diagram. Surprisingly, they 
did not detect any in the time-resolved radial velocity measurements for these 
stars, and showed that $3-5$ of the stars had magnetic field strengths 
considerably in excess of 2\,kG. While more evolved stars are theoretically 
expected to require relatively stronger magnetic fields to suppress local surface 
convection and facilitate observable amplitudes of rapid pulsations 
(\citealt{cunha02}), this study made a good case for the existence of noAp stars 
near the terminal age main sequence.

\subsection{HD\,151878}

HD\,151878A ($\alpha_{2000.0}= 16\,48\,41.5$, $\delta_{2000.0}= +35\,55\,19$, 
$V=7\fm22$) is the primary component of the visual binary HD\,151878 and is 
separated by 5\farcs8 from its 1.51\,mag fainter companion. \citet{bidelman85} 
used low resolution, objective prism plates and classified the star as Am:, where 
the colon means that the star is a marginal metallic line star with a difference 
between its spectral type determined from the Ca\,\textsc{ii} K\,line and from the 
metals lines of less than 5 spectral subclasses. \citet{grenier99} classified the 
star as A6mF0F5, where the A6 subtype refers to the Ca\,\textsc{ii} K\,line, F0 
refers to the H spectral type, and F5 refers to the metal line type. With 9 
spectral subclasses difference between the Ca\,\textsc{ii} K\,line type and metal 
line type, this classification indicates a strong classical Am star. Am stars are 
characterised by overabundance of iron-peak elements and underabundance of Ca or 
Sc. Importantly, the Ca deficiency of the Am stars is not a characteristic of the
Ap stars (see, e.g., \citealt{ryabchikovaetal00} and references therein for a study of
7 roAp stars), making it 
unlikely that this star is a magnetic Ap star. No medium or high dispersion 
spectroscopy has been published for the system prior to our study. The Str\"omgren 
indices of HD\,151878 (see Section~\ref{sec:params}) indicate spectral peculiarity 
and are consistent with either an Am or Ap spectral type. 

\citet{tiwarietal07} recently announced the discovery of rapid photometric 
oscillations in HD\,151878 with a period of 6\,min and an amplitude of as much as 
22\,mmag peak-to-peak, making this the highest amplitude roAp star known by a 
significant margin. The previous highest photometric amplitude observed for a roAp 
star was on the discovery night for HD\,60435 when a peak-to-peak variation of 
16\,mmag was detected for this star  (\citealt{kurtz84}). That star is the most 
multiperiodic of all roAp stars (\citealt{matthewsetal87}) and it rarely shows an 
amplitude this high. For roAp stars with a single, or dominant pulsation mode, the 
highest photometric $B$ amplitudes are around 10\,mmag peak-to-peak in, e.g., 
HD\,99563 (\citealt{handleretal06}), HD101065 (Przybylski's star; 
\citealt{martinez90}) and HD83368 (HR\,3831; \citealt{kurtzetal97}). 

Thus, the announcement by \citet{tiwarietal07} of high amplitude photometric 
pulsations in HD\,151878 was potentially very important for many reasons: (i) The 
photometric amplitude they found is the highest known for any rapidly oscillating 
A star; (ii) if the star is an Am star -- as two spectral classifications show -- 
then it would be the first known rapidly oscillating Am star; that would suggest 
that a strong magnetic field is not required for high overtone pulsation modes in 
chemically peculiar A stars, which would require a complete rethinking of 
theoretical models; (iii) if an error had been made in the spectral classification 
and HD\,151878 is an Ap star, then it would be only the 6th known roAp star in the 
northern hemisphere; and (iv) the revised Hipparcos parallax ($5.81\pm0.61$\,mas, 
\citealt{vanleeuwen07}), along with a spectroscopically estimated effective 
temperature of 7000\,K, shows that HD\,151878 has evolved past the terminal age main sequence
(Section\,\ref{sec:params}). The most evolved roAp star, HD\,116114, shows no 
photometric variability in Johnson $B$ above 1\,mmag on 6 nights of observation 
(\citealt{martinezetal94}), and has the lowest radial velocity amplitudes known 
for a roAp star. This supports theoretical expectations (\citealt{cunha02}) of 
systematically lower frequencies and amplitudes for more evolved roAp stars (at 
fixed magnetic field strength). HD\,151878 is inconsistent with this, unless it 
has a stronger field than that of HD\,116114 (6\,kG), which we show it does not.

Because of the importance of such high amplitude pulsation in HD\,151878, we 
obtained the first high resolution spectra of the star to study the star's 
pulsation characteristics, and to study its physical properties -- particularly 
its spectral peculiarities.

\section{Observations and data reduction}

A total of 74 spectra of HD\,151878 were obtained in 2.55\,h with the FIbre-fed 
Echelle Spectrograph (FIES) at the 2.56\,m Nordic Optical Telescope (NOT) on 2008 
April 30 (HJD\,$2454587.196-2454587.752$). The observations were collected by the 
observatory staff as part of the Fasttrack Service Mode programme of the NOT. The 
spectra cover the wavelength range $3860-7350$\,\AA\ at a resolution of 
$R\sim65000$. The aperture of the high resolution fibre used is 1\farcs3. With the 
46-s readout and overhead times of the fast mode of the CCD, and 80-s exposure 
times, we obtained a sampling rate of 29 spectra h$^{-1}$, i.e. 3 measurements 
per pulsation cycle for the photometric candidate frequency. The spectra were 
reduced with the {\sc FIEStool} software supplied by the 
NOT\footnote{http://www.not.iac.es/instruments/fies/fiestool/FIEStool.html}. This 
package, based on {\sc python} and {\sc PyRAF}  was especially developed for FIES and 
performs all conventional steps of echelle data reduction, including the 
subtraction of bias frames, modelling and subtraction of scattered light, 
flatfielding, order extraction, normalisation (including fringe correction) and 
wavelength calibration. Barycentric velocity correction was included in the 
wavelength calibration. The typical signal-to-noise ratio of the individual 
spectra is $40 - 60$ ($\lambda\lambda5500-6500$\,\AA), and about 350 in the 
combined spectrum.

A single ThAr reference spectrum, obtained immediately after the science spectra 
sequence, was found sufficient to do the Doppler shift test. Radial velocity 
measurements of telluric lines show a wavelength stability to better than 
100\,\ms\ over 2.55\,h, or to the accuracy obtainable from internal stability of 
telluric lines. The same stability is found with sharp lines in the stellar 
spectrum. Relative to $5-6$ telluric lines, the absolute wavelength scale's zero 
point deviates by only $-330\pm130$ (start) to $-200\pm170$\,\ms\ (end) over the 
2.55-h run. The stellar spectrum has a mean radial velocity of 
$-33.9\pm0.2$\,\kms.

The merged 1-D spectra were normalised relatively to a run master spectrum (a co-
added spectrum with a high signal-to-noise ratio), then divided by a detailed 
spline fit to the master spectrum, followed by a low-detail, low-frequency 
continuum fit performed to each individual spectrum. The line density is very high 
for HD\,151878, especially to the blue where there is no continuum, in spite of a 
sharp lined spectrum. Thus we focused our efforts on normalising the spectra 
for the \changeb{$5000-7000$\,\AA\ region} where continuum windows are present and where
our velocity and abundance analyses were performed. The region around the
Ca\,\textsc{ii}\,K line was additionally normalised 
by comparison to model spectra.
\begin{figure*}
\begin{center}
\hspace{-1cm}
\hfil \includegraphics[width=9.9cm, height=13.9cm, angle=270]{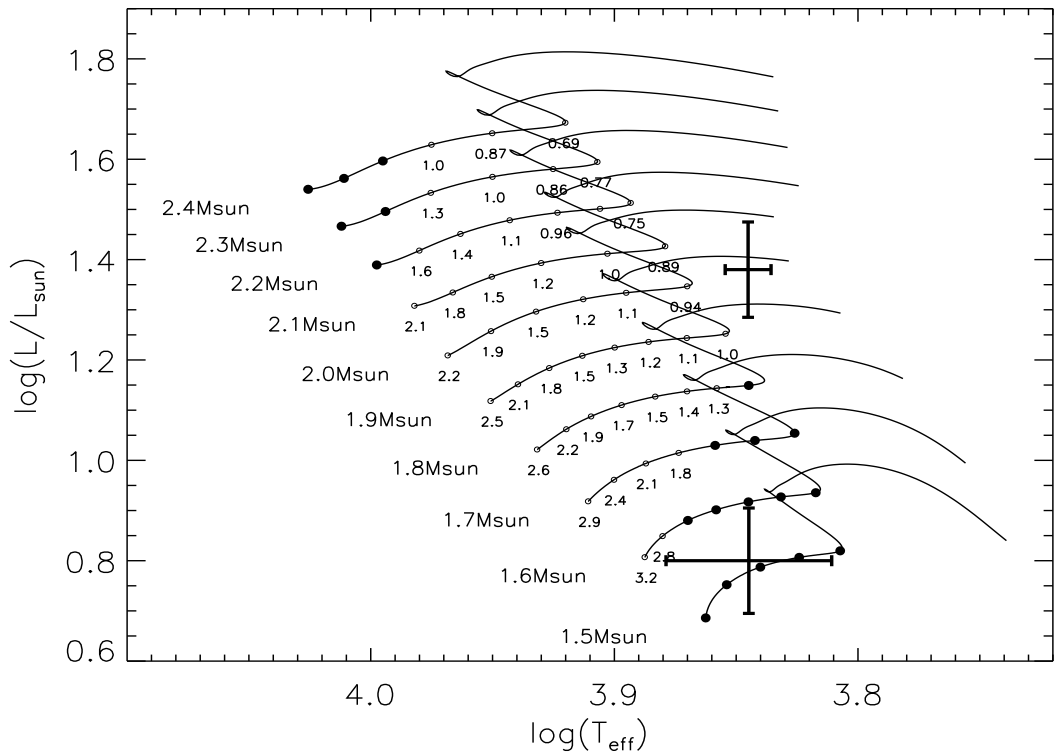}
\caption{\label{fig:hrd} Locations of the two components of HD\,151878 in the 
HR-diagram by \citet{cunha02}. Open circles indicate the predicted frequencies for 
pulsationally unstable rapid oscillation modes of roAp stars. Error bars indicate 
all known uncertainties for luminosities calculated
from the revised {\it Hipparcos} trigonometric parallax, the spectroscopic 
temperature estimate for the primary star, and an assumed \teff-range for the secondary. 
The evolved primary is located in the region typical for many $\rho$\,Pup stars 
(these are evolved Am stars).
Both components are outside the main sequence region of the roAp instability 
strip.}
\end{center}
\end{figure*}
\section{Spectroscopic analysis}
\subsection{The stellar parameters} \label{sec:stelpar}
\label{sec:params}
We first estimated the stellar parameters of HD\,151878 from Str\"omgren 
photometry compiled by \citet{haucketal97} and from \citet{olsen83}. From the 
$\beta$ index alone, $\beta=2.759$, we derived $T_{\rm eff} = 7300 \pm 150$\,K using 
the \citet{moonetal85} calibration. The Str\"omgren indices for 
the combined light of both components of HD\,151878 are: $V=7.223\pm0.005$, $b-
y=0.227\pm0.003$, $m_1=0.249\pm0.004$ and $c_1=0.677\pm0.004$. 
From the 
calibrations for A stars given by \changeb{\citet{crawford79}}, we derive $E(b-
y) = 0.040$, $\delta m_1 = -0.061$ and $\delta c_1 = -0.021$. These $\delta m_1$ 
and $\delta c_1$ indices indicate strong line blocking in the $u$ and $v$ filters 
and are typical of strongly peculiar Am and Ap star spectra. 
Using calibrations by 
\citet{napiwotzkietal93}, the {\sc TEMPLOGG} code\footnote{Available at 
http://ams.astro.univie.ac.at/templogg} indicates  that $E(b-y)=0.024$, $T_{\rm 
eff}=7100$, \logg=4.05 and $[$Fe/H$]=0.745$. We note that these calibrations are 
less reliable for stars with peculiar abundances and stratified atmospheres, than 
for normal stars.

However, a lower temperature is supported by the spectra. Model spectra were 
produced from Kurucz model atmospheres with $T_{\rm eff} = 7000, 7250, 7500$\,K 
and $\log g = 3.0$, 3.5, 4.0, solar metallicity ([M/H]$ =0.0$) and $\xi=2$\kms\ 
using {\sc SYNTH} \citep{piskunov92}. Spectral line lists were taken from the 
Vienna Atomic Line Database (VALD, \citealt{kupkaetal99}) and the DREAM database 
\citep{biemontetal99}, using a microturbulence of $\xi=2$\,\kms. The best fit to 
the wings of \halpha\ was obtained for the adopted temperature $T_{\rm eff} = 
7000\pm150$\,K and $\log g = 3.5\pm0.5$. With these parameters fixed, individual 
lines in the combined spectrum were directly compared to the synthetic line 
profiles to provide a first look at the spectroscopic nature of HD\,151878 
(Sect.\,\ref{sec:abund}).

{\it Hipparcos} obtained 138 measurements of the star that show no variability 
above 9.6\,mmag amplitude. 
A marginally significant 14.2\,h period is seen at that amplitude, but cannot be confirmed 
independently and does not agree well with the low \vsini\ we observe (see below), 
if related with the 
stellar rotation. The {\it Hipparcos} catalogue \citep{perrymanetal97} gives in 
its Double and Multiple Systems Annex individual apparent magnitudes for the two 
components of HD\,151878: $V_T({\rm A})=7.467\pm0.009$ and $V_T({\rm 
B})=8.924\pm0.022$. As the combined magnitude $V_T({\rm AB})=7.215$ agrees well 
with that of \citet{olsen83}, we here use the individual ones to characterise the 
two stars. It is assumed that the secondary still is on the main sequence, with a 
temperature of $T_{\rm eff}({\rm B})=7050\pm550$\,K. Using bolometric corrections 
from interpolation in the tables by \citet{flower96} (accounting for the 
uncertainty of $T_{\rm eff}$), the revised {\it Hipparcos} parallax $5.81\pm0.61$ 
mas \citep{vanleeuwen07} and a conservative interstellar absorption $A_V=0.050-
0.056$ mag from NASA's IR dust maps, provide (all errors included) $\log L_{\rm 
A}/{\rm L}_\odot = 1.38^{-0.09}_{+0.10}$ and $\log L_{\rm B}/{\rm L}_\odot = 
0.80^{-0.10}_{+0.11}$. 
Using the standard relation $L=4\pi\sigma R^2 T^4$ and the solar temperature 
$T_{\rm eff,\odot}=5785$\,K, then the relative radii of HD\,151878's stars 
become $R_{\rm A}/{\rm R}_\odot=3.35^{-0.34}_{+0.42}$ and $R_{\rm B}/{\rm 
R}_\odot=1.72^{-0.19}_{+0.24}$.

This places the primary star just past the H core 
exhaustion (Fig.\,\ref{fig:hrd}) with a mass of $1.82\pm0.06\; {\rm M}_\odot$. 
From the HR-diagram by \citet{cunha02}, rapid oscillations of $0.6 - 0.9$\,mHz are 
then predicted for HD\,151878's primary, in contrast with the 2.78\,mHz found by 
\citet{tiwarietal07}.

\subsubsection{Upper limit of a magnetic field}

The roAp star HD\,116114 has a 6-kG magnetic field that may be
important to sustain its pulsations as it is an evolved main sequence star. Being
more evolved, HD\,151878 is expected to have a similar, or 
stronger, magnetic field to exhibit observable rapid oscillations. Although strong magnetic
fields are not expected in Am stars, we used the FIES spectra to firmly exclude
such a strong field.  

\begin{figure}
\begin{center}
\hfil \epsfxsize 8.5cm\epsfbox{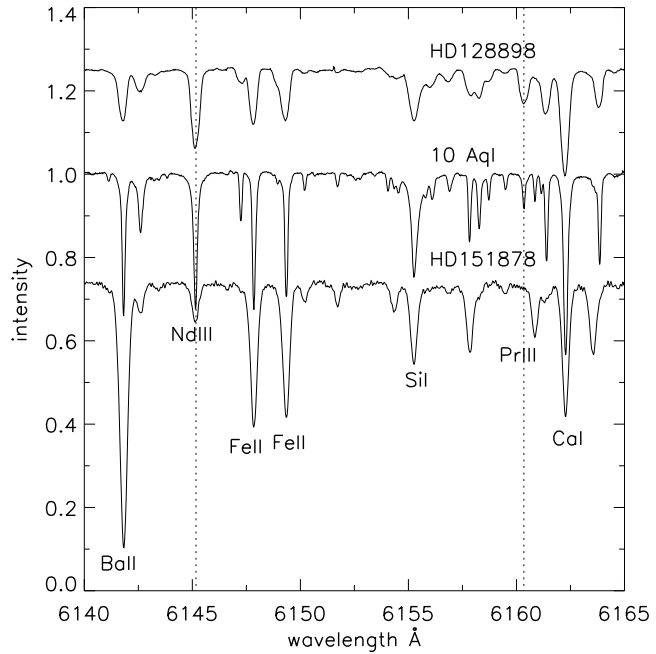}
\caption{\label{fig:reg1}A  normalised spectral region of HD\,151878 (bottom 
spectrum), compared with those of two roAp stars, HD\,128898 ($\alpha$\,Cir) 
and HD\,176232 (10\,Aql). The spectra are mutually shifted in intensity by 0.25 
for viewing. Selected atomic lines are indicated: dotted lines mark locations of 
the rare earth elements Nd and Pr. Note the relatively weak Nd and Pr lines of 
HD\,151878, for which Nd\,\textsc{iii}\,6145\,\AA\ is a blend totally dominated by
Si\,\textsc{i}\,6145\,\AA.}
\end{center}
\end{figure}

The magnetically 
sensitive line Fe\,\textsc{ii}\,6149\,\AA\ line (see Fig.\,\ref{fig:reg1}) 
is often used as a diagnostic line for determining the magnetic field 
modulus $\left < B\right >$ from the line splitting of its Zeeman components (see, 
e.g., \citealt{mathys89}; \citealt{freyhammeretal08b}). By comparing the
averaged FIES spectrum to model spectra calculated using {\sc SYNTHMAG} 
\citep{piskunov99} and rotationally broadened to $v \sin i=8$\,\kms, it
is seen that partial splitting would be visible in Fe\,\textsc{ii}\,6149\,\AA\ 
at 6.4\,kG. Already at 5.7\,kG, the line shape is very broad
with at flat core, highly different from the observed line. 

No magnetically resolved lines were found in the spectrum, and for 21 Fe lines
with different Land\'e factors (and magnetic sensitivity), no clear relation
of magnetic broadening with Land\'e factor could be detected for a
relatively high scatter of $\sigma(\left < B \right>)=2.0$\,kG. 
The line broadening was determined by fitting line widths with
Gaussian profiles.
The observed spectra therefore firmly exclude any field stronger 
than $\left < B \right> \sim 5.5$\,kG.

\section{Search for rapid radial velocity variations}

We searched for pulsations as periodic Doppler shifts in two ways: first by using 
cross correlations of long stretches of spectral regions, then by measuring 
centre-of-gravity shifts of individual lines in the spectra. For frequency 
analyses, we used a discrete Fourier transform programme by \citet{kurtz85} and 
the {\sc PERIOD04} \citep{lenzetal05} programme. For the run length of 
2.55\,h and sampling interval of 126\,s, the frequency resolution is about 
0.1\,mHz and the Nyquist frequency 4.0\,mHz. Without simultaneous reference 
spectra, we have little control over spectrograph stability on time scales of 
about an hour, so we only consider the frequency range $0.4 - 4.0$\,mHz.

\subsection{Cross correlation radial velocity analyses}

The cross correlation method, using large spectral regions, has been successful 
for detecting pulsation in roAp star candidates and for finding additional 
frequencies in known roAp stars (see, e.g., \citealt{matthewsetal88}, 
\citealt{balonaetal02}). The cross correlation amplitudes from correlation of long 
spectral regions are, though, not directly comparable to those derived from line 
profile measurements. This is mainly due to the different pulsation amplitudes of 
different elements in the stratified roAp atmospheres, where low amplitude 
elements such as Fe dilute the `integrated' Doppler shifts. However, for detecting 
low amplitude pulsations, the method is very efficient, especially in cases where 
amplitude and phase is comparable for the most abundant elements.

\begin{figure}
\begin{center}
\hfil \includegraphics[height=8.3cm, angle=270]{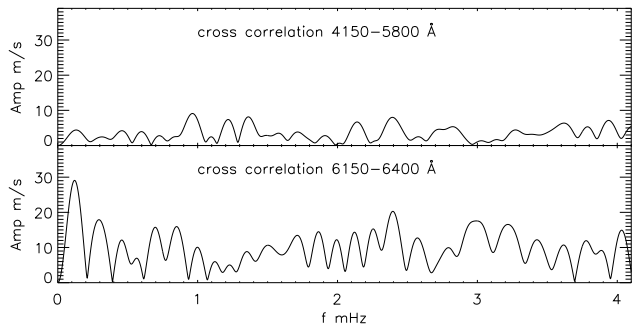}
\caption{\label{fig:xcor} Amplitude spectra for cross correlations of two 
indicated wavelength regions of HD\,151878. There are no significant peaks, 
notably, in this case, at the claimed photometric frequency of 2.78\,mHz. }
\end{center}
\end{figure}

Cross correlations were performed with our HD\,151878 spectra, using the average 
spectrum as template. For the line-rich spectral regions $\lambda\lambda4150-5800$ 
and $5150-5800$\,\AA, no variability is seen above 10\,\ms\ ($\sigma=3-4$\,\ms, 
see Fig.\,\ref{fig:xcor}) in the frequency domain of known roAp stars. The 
wavelength region $\lambda\lambda6150-6400$\,\AA\ has a lower line density and a 
higher scatter, but excludes variability above 20\,\ms\ ($\sigma=7$\,\ms).

The spectral region $\lambda\lambda6863 - 6938$\,\AA\ with abundant telluric lines 
was used to check the instrumental stability and identify non-stellar 
periodicities. The telluric lines only showed low frequency noise $<0.4$\,mHz, due 
to instrumental drifts or meteorological changes: a mostly linearly decreasing 
velocity of $-95$\,\ms, and elsewhere stability at the level below 9\,\ms\ 
($\sigma=3\,\ms$).

%--------------------------------------------------------------------------
\subsection{Line profile radial velocity analyses}

\begin{figure*}
\begin{center}
\hspace{-3cm}
\epsfxsize 13.0cm\epsfbox{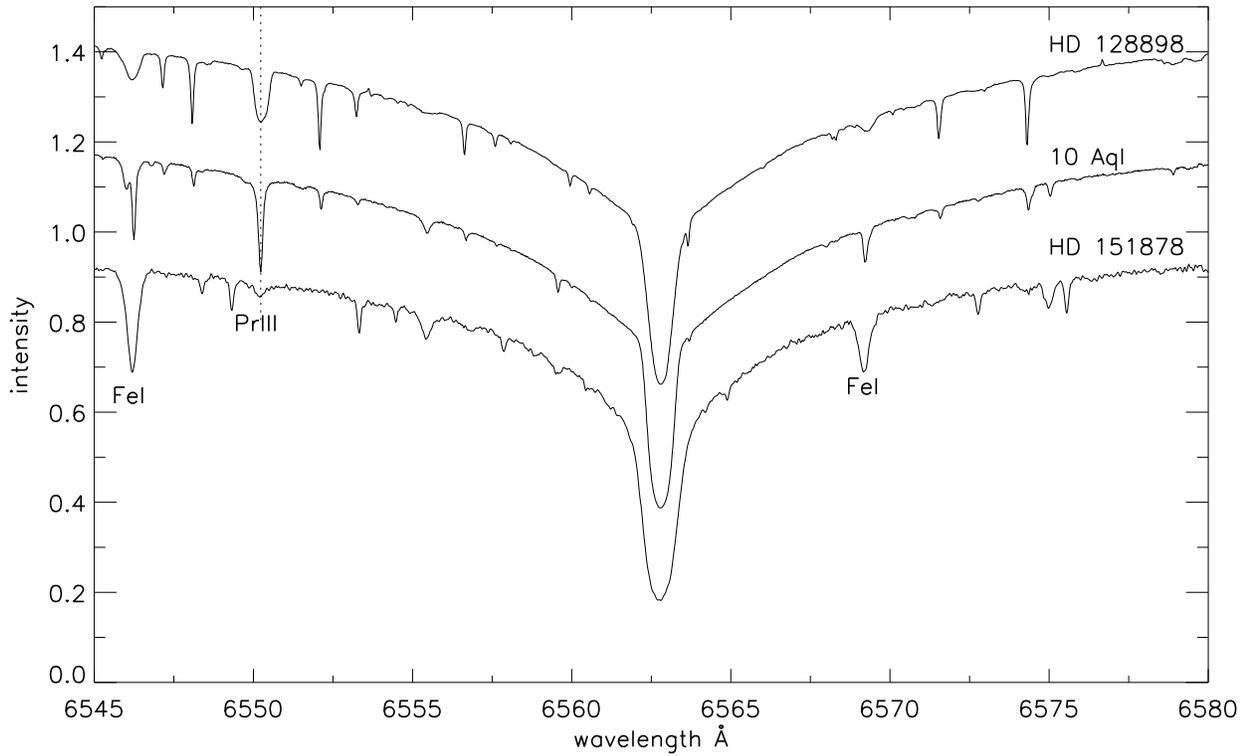}
\hfil
\caption{\label{fig:ha}Same as Fig.\,\ref{fig:reg1}, but for the spectral region 
in the vicinity of the \halpha\ line, again compared with those of the two 
roAp stars. Note that the -- for roAp stars typical -- core-wing anomaly 
\citep{cowleyetal01} is much less pronounced for HD\,151878.
}
\end{center}
\end{figure*}

\begin{figure*}
\begin{center}
\hfil \includegraphics[width=10.5cm, height=13.5cm, angle=270]{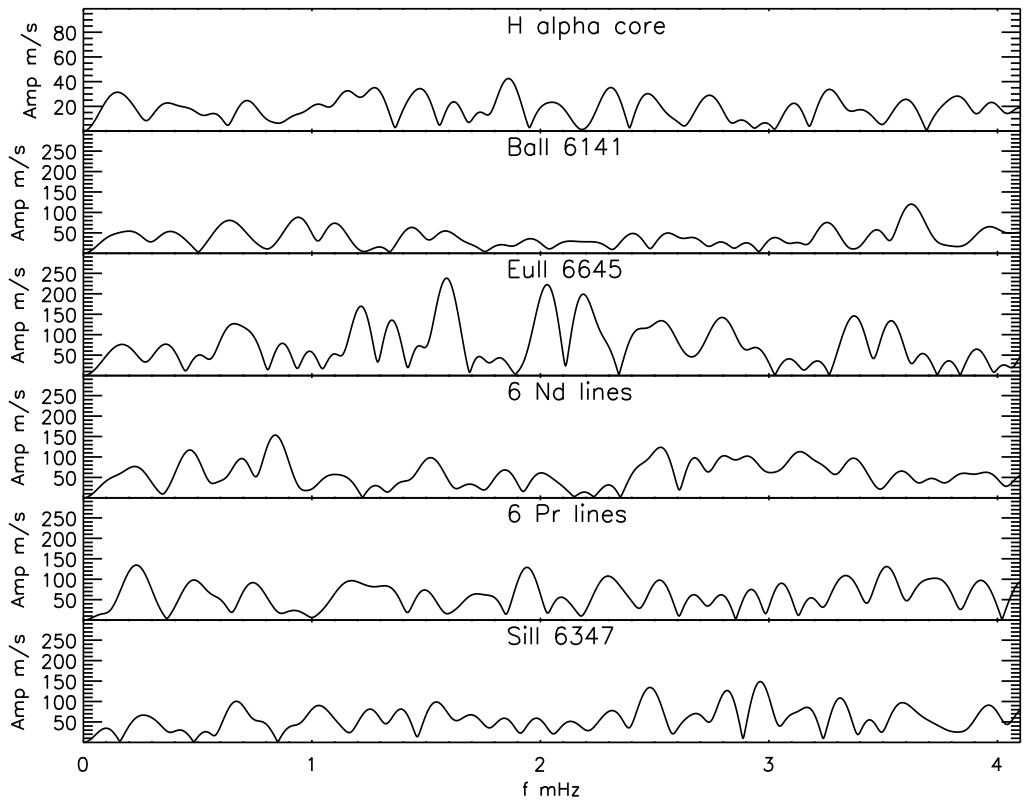}
\caption{\label{fig:cog} Amplitude spectra for individual and combined RV time 
series for selected elements. Note the different ordinate scales. Eu shows some 
power in the signal around 2\,mHz, but the dominant peaks are not in common with 
peaks for the other two studied Eu lines. }
\end{center}
\end{figure*}

In roAp stars, lines of rare earth elements typically show the largest Doppler 
shift pulsation amplitudes. Amplitudes vary for different elements and range from 
a few dozen metres per second up to a few kilometres per second for various roAp 
stars \citep{elkinetal08a}. Also, the narrow line core of the H$\alpha$ profile 
(see Fig.\,\ref{fig:ha}) shows rapid pulsations in roAp stars 
\citep{kurtzetal06,elkinetal08a}. We therefore searched for pulsations in 
HD\,151878 using such lines and show amplitude spectra of a subset of these in 
Fig.\,\ref{fig:cog}. Although we concentrate on analyses of lines of the rare 
earths, other chemical elements were also tested. All radial velocity curves 
subjected to period searches were first de-trended with linear least-squares 
fitting.

The H$\alpha$ core was stable with highest peaks in the amplitude spectrum of 
43\,m\,s$^{-1}$ with an amplitude standard deviation noise level of 13\,m\,s$^{-
1}$; H$\beta$ was stable to 85\,\ms\ ($\sigma=22$\,\ms). Of the rare earth 
elements, lines of \changeb{Ce\,\textsc{ii}, Pr\,\textsc{ii}, Pr\,\textsc{iii}, 
Nd\,\textsc{ii}, Nd\,\textsc{iii}}, Eu\,\textsc{ii}, Tm\,\textsc{ii} and 
La\,\textsc{ii} excluded rapid pulsations with typical upper limits of 
320\,m\,s$^{-1}$ and noise levels varying from 70 to 110\,m\,s$^{-1}$ for the 
majority of good lines. Combining lines of the same ion reduced the noise level, 
such as 6 Nd lines: $A_{\rm max}=153$\,\ms\ ($\sigma=45$\,\ms) and 6 Pr lines: 
$A_{\rm max}=135$\,\ms\ ($\sigma=42$\,\ms), or combined for all 12 lines $A_{\rm 
max}=111$\,\ms\ ($\sigma=31$\,\ms). 

Of the non-rare earth element lines analysed, including the strong sodium D and 
Mg\,\textsc{i} lines and several Ca, Sc, Ti, Cr, Fe, Y and Ba lines, no pulsations 
were detected above typical upper limits of $100-200$\,m\,s$^{-1}$ ($\sigma=30-
100$\,\ms). In general, some lines, such as Eu in Fig.\,\ref{fig:cog} show 
intriguing peaks in the amplitude spectra for the frequency range typical for roAp 
stars. But as these peaks could not be confirmed by other spectral lines of the 
same element or by other rare earth lines, they were rejected as indications of 
pulsation in the star (using a 4$\sigma$ significance criterion). Iron is strong 
and abundant in HD\,151878 but normally exhibits the lowest pulsation amplitudes 
in known roAp stars. With 15 lines, the typical upper amplitude limit for 
HD\,151878 was 150\,\ms\ ($\sigma=40-45$\,\ms).

We conclude that there is no evidence of pulsation in the radial velocities of our 
spectroscopic times series. 

%--------------------------------------------------------------------------
\subsection{Chemical abundances}
\label{sec:abund}

The presence of high overabundances of rare earth elements is one of the 
characteristics of known roAp stars. Another important property is that the 
abundances of the first two ionised states of neodymium and praseodymium show more 
than 1\,dex difference, with the doubly ionised ions, which form higher in the 
atmosphere, giving the highest abundance \citep{ryabchikovaetal04,kurtzetal07}. 
This ionisation disequilibrium anomaly may be explained mostly by concentration of 
rare earth elements in high atmospheric layers (stratification) and partly by 
\changeb{non-LTE} 
effects \citep{mashonkinaetal05}. \changeb{Non-LTE} effects may enhance the 
ionisation of Nd\,\textsc{ii} or Pr\,\textsc{ii} and accordingly weaken their 
absorption lines, while strengthening those of the second ionisation state. 

To compare HD\,151878 spectrally with the roAp stars, we estimated its chemical 
abundances for selected elements and tested for ionisation disequilibria for Nd 
and Pr. Abundances were determined by fitting synthetic spectra to the observed 
average spectrum. The model spectra were calculated with {\sc SYNTH} as described in 
Sect.\,\ref{sec:stelpar}, using a Kurucz model atmosphere with $T_{\rm eff} = 
7000$\,K, $\log g = 3.5$ and slightly enhanced (above solar) metallicity:
 [M/H]$ = 
          \log ({\rm N}_{\rm M}/{\rm N}_{\rm H}) -
          \log ({\rm N}_{\rm M}/{\rm N}_{\rm H})_\odot =+0.2$.
We tried microturbulence values of $\xi=2$, 4 and 5\,\kms\ and
found the best agreement in derived abundances for strong and weak iron lines using  
$\xi=4$\,\kms, which was then adopted.  A projected rotation velocity of
$v \sin i = 8.3\pm0.6$\,km\,s$^{-1}$ was determined  from several symmetric iron lines, 
and the macroturbulence velocity was optimised, on a line-by-line basis, in the 
range $5 - 8$\,\kms. The latter parameter was chosen such that
the model spectra matched wings and core of the observed line profiles. However,
in some cases could a 1--2\,$\sigma$ larger \vsini\ also be used for a lower
macroturbulence velocity.

The resulting abundances for the 2.55\,h average spectrum are presented in 
Table\,\ref{tab:abund} and shown graphically in Fig.\,\ref{fig:ab}. Solar abundances 
are from \citet{asplundetal05}. Relative abundances are given on the scale 
     $\log\,\epsilon_{\rm El}-\log\,\epsilon_{\rm tot,}\odot=
          \log ({\rm N}_{\rm El}/{\rm N}_{\rm tot}) -
          \log ({\rm N}_{\rm El}/{\rm N}_{\rm tot})_\odot$.
Most of the measured elements are stronger than solar, increasing in abundances
with atomic number Z. Lines of Ca and Sc\,\textsc{ii} have significantly less than 
solar abundances. Ca\,\textsc{i} and Ca\,\textsc{ii} lines were all deficient 
(including the Ca\,\textsc{ii} H and K lines) and give combined  $\log \epsilon
{\rm (Ca)}=5.59\pm0.14$\,dex. Ca and Sc deficiency and overabundance of metallic
elements are Am star characteristics. 
The Ca deficiency is consistent 
with the spectral classification of \citet{grenier99}, which suggests strongly 
deficient Ca\,\textsc{ii} from the weakness of the K line, but is inconsistent with 
the Am: (marginal Am) classification of \citet{bidelman85}. The former 
classification is, however, more consistent with the Str\"omgren $\delta m_1$ and 
$\delta c_1$ indices. Figure\,\ref{fig:CaK} shows our observed Ca\,\textsc{ii}\,K line
compared to model spectra for 3 different Ca abundances. A lower than solar 
Ca\,\textsc{ii} abundance is required, which confirms Grenier's result and 
the Am status of the star.

\begin{figure} 
\begin{center} 
 \includegraphics[height=8.2cm, angle=90]{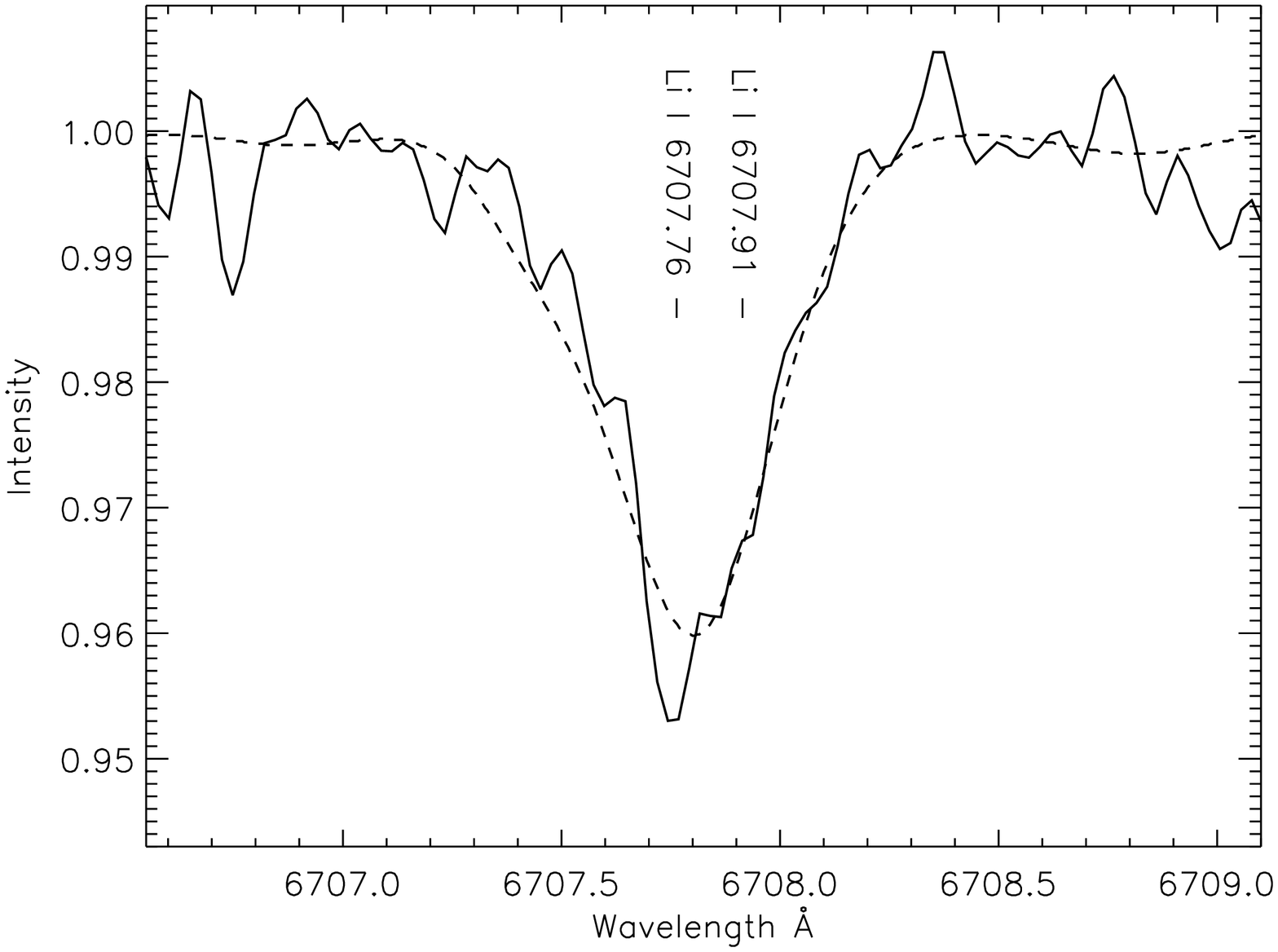}
\caption{\label{fig:li} The Li blend Li\,\textsc{i}\,6707.76 and 
Li\,\textsc{i}\,6707.91\,\AA\ with a rotationally broadened 
model spectrum with $\log\,\epsilon{\rm (Li)}=2.74$ dex.
} 
\end{center} 
\end{figure}
\begin{figure} 
\begin{center} 
 \includegraphics[height=8.4cm, angle=90]{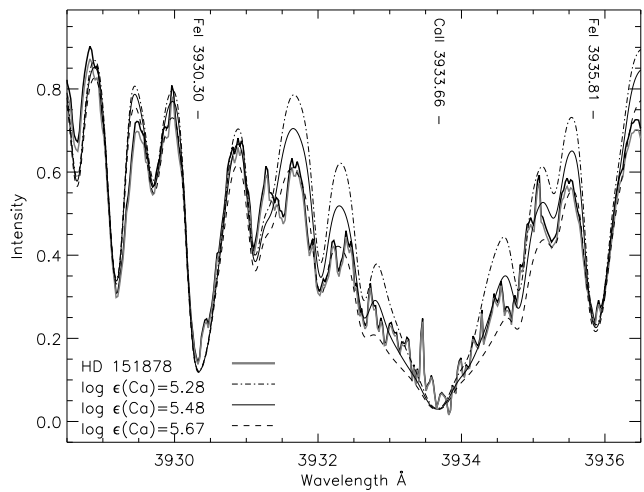}
\caption{\label{fig:CaK} The observed Ca\,\textsc{ii}\,K\,3933.66\,\AA\ line
(thick grey line), compared with model spectra for three labelled 
Ca abundances. The model spectra were broadened to \vsini=8\,\kms.
} 
\end{center} 
\end{figure}

We do not find indication of ionisation disequilibria for neither neodymium or 
praseodymium and simply give the combined abundance of the first two ionized 
states in Table\,\ref{tab:abund}. 
The abundances for Nd, Pr and Eu are about 1\,dex above solar, which is low for 
most known roAp stars, such as HD\,128898 and 
10\,Aql (Figs.\,\ref{fig:reg1} and \ref{fig:ha}). Thus, we find HD\,151878 is spectrally 
comparable to Am stars, and less to Ap stars.
A minor ionisation disequilibrium for Cr was noted:
$\log\,\epsilon_{\rm Cr I}=5.92\pm0.09$ and 
$\log\,\epsilon_{\rm Cr II}=6.07\pm0.14$.
Li (Fig.\,\ref{fig:li}) is the most overabundant of the studied
elements (Fig.\,\ref{fig:ab}). \citet{northetal05} found that while all
Li-deficient Am stars appear to be evolved, not all evolved Am stars are Li-deficient.
HD\,151878 supports the latter statement.

\begin{table} \caption[] {\label{Table1}Chemical abundances for HD\,151878 for 
selected elements, and their corresponding solar abundances \citep{asplundetal05}. 
The errors quoted are internal standard deviations for the set of lines measured.
The two measured Li lines are components
in the same blend, Li\,\textsc{i}\,6707.76 and Li\,\textsc{i}\,6707.91\,\AA.
}
\begin{center}
\begin{tabular}{lcrr}
\hline
\multicolumn{1}{c}{Ion} & \multicolumn{1}{c}{Number} &
\multicolumn{1}{c}{$\log\,\epsilon$} & \multicolumn{1}{c}{$\log\,\epsilon$}
\\
& \multicolumn{1}{c} {of lines} & \multicolumn{1}{c}{HD\,151878} &
\multicolumn{1}{c}{Sun} \\ \hline
Li             &2   & $2.69\pm0.20$ & $1.05\pm0.10$ \\  
Na             &3   & $6.52\pm0.09$ & $6.17\pm0.04$ \\
Si             &7   & $7.79\pm0.06$ & $7.51\pm0.04$ \\
S              &3   & $7.48\pm0.15$ & $7.14\pm0.05$ \\
Ca\,\textsc{i} &5   & $5.71\pm0.06$ & $6.31\pm0.04$ \\
Ca\,\textsc{ii}&4   & $5.48\pm0.20$ & $6.31\pm0.04$ \\
Sc             &4   & $1.13\pm0.40$ & $3.05\pm0.08$ \\
Ti             &4   & $4.58\pm0.31$ & $4.90\pm0.06$ \\
Cr             &16  & $5.98\pm0.14$ & $5.64\pm0.10$ \\
Fe             &14  & $7.63\pm0.07$ & $7.45\pm0.05$ \\
Co             &5   & $5.33\pm0.06$ & $4.92\pm0.08$ \\
Ni             &12  & $6.83\pm0.09$ & $6.23\pm0.04$ \\
Cu             &4   & $4.71\pm0.07$ & $4.21\pm0.04$ \\
Y              &7   & $3.04\pm0.16$ & $2.21\pm0.02$ \\
Zr             &3   & $3.13\pm0.20$ & $2.59\pm0.04$ \\
Ba             &3   & $3.37\pm0.06$ & $2.17\pm0.07$ \\
La             &3   & $2.36\pm0.10$ & $1.13\pm0.05$ \\
Ce             &4   & $2.73\pm0.10$ & $1.58\pm0.09$ \\
Pr             &11  & $1.72\pm0.28$ & $0.71\pm0.08$ \\
Nd             &10  & $2.35\pm0.20$ & $1.45\pm0.05$ \\
Eu             &3   & $1.62\pm0.12$ & $0.52\pm0.06$ \\ \hline \hline
\end{tabular}
\label{tab:abund}
\end{center}
\end{table}

\begin{figure*}
\begin{center}
\includegraphics[width=9.0cm, height=14.0cm, angle=90]{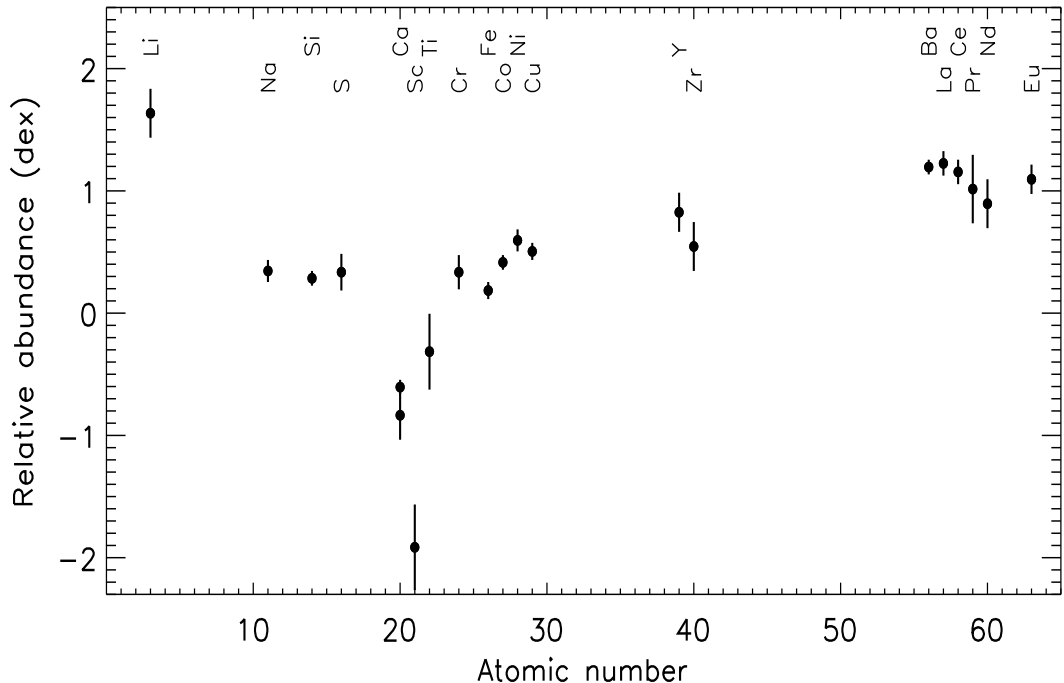}
\caption{\label{fig:ab} Relative abundances ($\log\,\epsilon-\log\,\epsilon_\odot$) 
for selected elements in the photosphere of HD\,151878. Error bars indicate the 
1$\sigma$ scatter among the measured lines. Ca abundances for the two first ions 
are indicated, the lowest (less than solar) being for Ca\,\textsc{ii}.}
\end{center}
\end{figure*}

\section{Discussion and summary}

The known roAp stars have radial velocity amplitudes in the range of few \ms\ to 
over 5\,\kms\ for some rare earth element lines. Low amplitude oscillations in roAp 
stars are known from, e.g., $\beta$\,CrB \citep{kurtzetal07}, HD\,154708 
\citep{kurtzetal06} and HD\,116114 \citep{elkinetal05a}, the latter with 
amplitudes of $26-125$\,\changeb{\ms}, measured in lines of the rare earth elements Eu, La 
and Zr and in the core of the H$\alpha$ line. The highest radial velocity 
amplitudes yet detected are for lines of Eu\,\textsc{ii} in HD\,99563 
(\citealt{elkinetal05b}) where individual components generated by surface spots 
can reach amplitudes of even 8\,km\,s$^{-1}$. The roAp stars also have strong 
magnetic fields that range from several hundred Gauss to 24.5\,kG in HD\,154708. 
Furthermore, they have strong abundances of rare earth elements and almost all of 
them show ionisation disequilibria for Pr and Nd.

HD\,151878 fails to show these characteristics and is spectroscopically unlike the  
roAp stars. It shows no rapid oscillations with radial velocity amplitudes above a 
few tens of m\,s$^{-1}$ at the time of our observations. The methods applied, and 
the length of the data set acquired, would be sufficient for detecting pulsations 
in most known roAp stars. The roAp stars are oblique pulsators with amplitude 
modulation of their pulsation with rotation. There is no clear correlation between 
photometric pulsation amplitude and radial velocity pulsation amplitude in the 
roAp stars, but the higher photometric amplitude roAp stars -- e.g. HD\,83368, 
HD\,60435, HD\,99563, HD\,101065 -- have H$\alpha$ radial velocity amplitudes of 
hundreds of m\,s$^{-1}$ to 2.5\,km\,s$^{-1}$; we would expect HD\,151878 to be 
similar. Therefore, for the spectacularly high photometric amplitude reported by 
\citet{tiwarietal07}, it is improbable that we observed at a rotation phase where 
the radial velocity amplitude was undetectable at the high precision of our 
measurements. 

The evolutionary tracks in Fig.~\ref{fig:hrd}, as well as those in figure 1 of 
\citet{cunha02}, indicate that HD\,151878 is past the terminal age main sequence. 
\citet{cunha02} argued that more evolved roAp stars require stronger magnetic 
fields to suppress the upper envelope convection and enable the rapid oscillations 
to reach observable amplitudes. She further speculated that the magnetic field 
intensities needed for suppressing the convection in roAp stars more often are 
found in roAp stars with magnetically resolved lines. That typically requires a 
magnetic field modulus of $\sim3$\,kG for slowly rotating Ap stars ($v \sin i=1-
3$\,\kms). 

\citet{freyhammeretal08a} searched for pulsations in 9 evolved cool Ap stars 
located inside the theoretical instability strip, the majority having estimated 
temperatures below $8100$\,K. With only slightly better precision and upper limits 
on radial velocity amplitudes as in this study, these authors found 9 null 
results. Out of 7 stars, only 3 stars had magnetic fields significantly stronger 
than 2\,kG, which possibly explains most of their null results for such evolved 
stars. 

Based on comparison of synthetic spectra 
with HD\,151878's Fe\,\textsc{ii}\,6149\,\AA\ line, we exclude 
magnetic fields stronger than $5.5$\,kG in this star. 
The magnetic field seems too weak for the star's evolved 
state to suppress local convection and enable pulsational driving of rapid 
pulsations to reach observable amplitudes. In this respect, the small, or absent, 
magnetic field is another Am star characteristic.

Our abundance analysis found that HD\,151878 has only mildly strong abundances of 
rare earth elements, no ionisation disequilibria for Nd and Pr and inhomogeneous 
surface distributions of these elements. In all these respects HD\,151878 does not 
resemble known roAp stars. Its Ca\,\textsc{ii} and Sc deficiencies is a clear sign 
of an Am, rather than 
Ap, nature. Our abundance analysis agrees with the Am spectral classifications. 
We conclude that HD\,151878 is a cool evolved Am star. As a such, rapid 
oscillations are unexpected. 

Evolved Am stars are known to be potential pulsators (unlike their main sequence 
counterparts, which rarely are), exhibiting $\delta$\,Sct pulsations (such as 
$\delta$\,Delphini). In fact, the evolutionary state and dimensions of HD\,151878 
resemble that of the pulsating evolved cool Am star HD\,98851 \citep{joshietal03}. 
However, the time scales of pulsating Am stars are much longer ($\sim$ hours) than 
those of roAp stars ($\sim$ minutes). Our spectra only cover 2.55\,h and are not 
suited for searching for $\delta$\,Sct pulsations. From our velocity measurements 
of \halpha\ and H$\beta$ we see no indication of variability on timescales of $1-
3$\,h and amplitudes above $60-250$\,\ms\ (without removing linear trends).

Could HD\,151878B be the source of the pulsations? This star was inside the 
diaphragm of the photometric observations of \citet{tiwarietal07}, but to cause a 
22\,mmag change in the combined light, the much fainter secondary would have to 
pulsate with an amplitude of 100\,mmag. While such an amplitude is characteristic 
of $\delta$\,Scuti stars, it is unheard of a roAp star (the highest known is 
16\,mmag for HD\,60435). The secondary star is located at the red border of the 
roAp instability strip, and is even possibly outside of it. We conclude that the 
secondary star could not be the source of the observed 6-min, 22-mmag photometric 
variations. 

This leads to the obvious question as to what is the cause of the observed 
variations reported by \citet{tiwarietal07}. Their light curves are convincing; 
the peaks in their amplitude spectra are highly significant. Their observations 
were obtained with the 1-m Sampurnanand telescope of \textsc{ARIES} at Naini Tal, 
India. The same telescope has been used for many years in the `Naini Tal -- Cape 
survey' for northern roAp stars (see, e.g., \citealt{joshietal06}, and references 
therein). Joshi et al. discovered apparent roAp pulsations in another star 
classified as an Am star, HD\,207561 (see their figure 3 and accompanying 
discussion). On two consecutive nights it showed significant high amplitude 
pulsation with a frequency of 2.75\,mHz ($P = 6.1$\,min) -- {\it the same frequency 
as found by \citet{tiwarietal07} for HD\,151878} -- yet no further signal was seen 
again in observations obtained on 17 individual nights over three observing 
seasons. We therefore suggest that the signal published for HD\,151878 is 
instrumental in origin. 

\section{Acknowledgements}  We thank Margarida Cunha for the evolutionary tracks 
and other data for constructing Fig.\,1 of this paper, which is directly based on 
figure 1 of \citet{cunha02}. \changeb{An anonymous referee is thanked for his/her
comments}. This research has made use of the NASA/IPAC Infrared 
Science Archive, which is operated by the Jet Propulsion Laboratory, California 
Institute of Technology, under contract with the National Aeronautics and Space 
Administration. We acknowledge using extracts of the introduction by 
\citet{elkinetal08b}. This research has made use of data obtained using the UK's 
AstroGrid Virtual Observatory Project, which is funded by the Science \& 
Technology Facilities Council and through the EU's Framework 6 programme. We
acknowledge support for this work from the Particle Physics and Astronomy 
Research Council (PPARC) and from the Science and Technology Facilities Council 
(STFC) and thank the NOT telescope staff for advice in planning and for
collecting the observations.

{}

\end{document}